\documentclass[%
 reprint,
 amsmath,amssymb,
 aps,
 prl,
 twocolumn
]{revtex4-1}

\usepackage{graphicx}
\usepackage{dcolumn}
\usepackage{bm}
\usepackage[qm,braket]{qcircuit}
\usepackage{slashed}
\usepackage[usenames, dvipsnames]{color}
\usepackage{hyperref}
\usepackage{csquotes}
\usepackage{relsize}

\newcommand\beq{\begin{eqnarray}}
\newcommand\eeq{\end{eqnarray}}

\newcommand\fig[1]{Fig.~(\ref{fig:#1})}

\newcommand\braket[2]{\langle #1 | #2 \rangle}

\usepackage[section]{placeins}
\newcommand{\ig}{\includegraphics}

\begin{document}

\preprint{INT-PUB-17-039}

\title{Ground States via Spectral Combing on a Quantum Computer}
\author{David B. Kaplan}
\email{dbkaplan@uw.edu}
\affiliation{%
Institute for Nuclear Theory, Box 351550, University of Washington, Seattle, WA 98195-1550
}%
\author{Natalie Klco}
\email{klcon@uw.edu}
\affiliation{%
Institute for Nuclear Theory, Box 351550, University of Washington, Seattle, WA 98195-1550
}%
\author{Alessandro Roggero}
\email{roggero@uw.edu}
\affiliation{%
Institute for Nuclear Theory, Box 351550, University of Washington, Seattle, WA 98195-1550
}%

\date{\today}

\begin{abstract}
A new method is proposed for determining the ground state wave function of a quantum many-body system on a quantum computer, without requiring an initial trial wave function that has good overlap with the true ground state.  The technique of Spectral Combing involves entangling an arbitrary initial wave function with a set of auxiliary qubits governed by a time dependent Hamiltonian, resonantly transferring  energy out of the initial state through a plethora of avoided level crossings into the auxiliary system. The number of avoided  level crossings grows exponentially with the number of qubits required to represent the Hamiltonian, so that the efficiency of the algorithm does not rely on any particular energy gap being large. We give an explicit construction of the quantum gates required for the realization of this procedure and explore the results of classical simulations of the algorithm on a small quantum computer with up to 8 qubits. We show that for certain systems and comparable results, Spectral Combing  requires fewer quantum gates to realize than  the Quantum Adiabatic Algorithm.
\end{abstract}


\maketitle

\section{Introduction}
One of the triumphs of physics and chemistry in the past century is the discovery of detailed theories for the microscopic interactions of the constituents of matter. These theories should in principle give us precise predictions for the properties of the nuclear matter in neutron stars, of large molecules, and  of interesting synthetic materials.  However, the complexity of such many-body quantum mechanical systems requires that they be studied largely numerically and, because Hilbert spaces grow exponentially with the number of constituents, only modest progress has been possible in these directions.  The cases where computation on a classical computer are successful include those where the ground state has special symmetry such as charge conjugation,  or where the quantum wave function can be well approximated by a product of single particle wave functions in background classical fields.
However, often a good representation of the wave function requires a number of computations that grows with the dimension of the Hilbert space, which itself grows exponentially with the size of the system \cite{booth2009fermion}.
In most formulations, this appears as a ``sign problem" where one must consider many configurations whose contributions to the final result exhibit strong cancellations \cite{Loh:1990zz,Troyer:170201,Muroya:2003qs,ortiz2001quantum,deForcrand:2010ys}.

Ground states are of special interest in nature because interesting systems are often in contact with a cold environment.  It has been shown that algorithms for finding the ground states of many-body systems on a quantum computer, in some cases, can scale with a power of the number of constituents rather than exponentially.  The improvement comes from being able to explore a Hilbert space of size $2^N$ with $N$ qubits.   Interesting examples of such algorithms include  Quantum Phase Estimation (QPE) \cite{nielsen2010quantum} and its variations, such as RFPE \cite{wiebe2016efficient}.  With QPE, one can use measurements to  project a trial wave function $\ket{\psi}$ onto those eigenstates of ${\mathcal H}$ with which it has significant overlap.   In this manner, the ground state wave function can be obtained precisely through rejection sampling with costs that scale as $\lvert\langle\psi_\text{gs}\ket{\psi}\rvert^{-2}$.
Another procedure is the Quantum Adiabatic Algorithm (QAA) \cite{farhi2000quantum,farhi2001quantum}, where the exact ground state of a simple initial Hamiltonian is adiabatically evolved into the ground state of the final Hamiltonian of interest.  Each approach has its limitations.  QPE requires  a good initial guess for the ground state, which may itself be a computationally hard problem for a strongly correlated system. The QAA requires maintaining adiabaticity and avoiding a Landau-Zener transition \footnote{For systems with well-understood interpolation spectra, multiple Landau-Zener transitions may be cleverly traversed to create an algorithmic advantage}; this is always theoretically possible if the time evolution of the interpolating Hamiltonian is sufficiently slow, but the computational effort scales  inversely with the square of the minimum value for the ground state energy gap, which could be extremely small in interesting cases.

Examples of physical systems that are anticipated to be difficult to analyze on a quantum computer with these existing methods  include: the unitary fermion gas in an external magnetic field, the Hubbard model away from half filling, QCD at finite baryon density, and systems exhibiting frustration such as spin glasses,  random Ising models, and the MAX k-SAT problem \cite{PapadimitriouCC,TrainingQoptimizer,Max2SATqubits}.  Systems with ground states that exhibit complex correlations are also likely to be difficult, such as nuclear matter near the liquid-gas phase transition \cite{Finn1982,BERTSCH19839}, large open-shell systems (both atomic and nuclear), systems coupled to random time dependent external fields, and those exhibiting the Anderson Catastrophe \cite{anderson1967infrared} due to the presence of sharp Fermi surfaces and non-perturbative impurities.
Given the diversity and complexity of ground state entanglement  expected in interesting physical systems, it is unlikely that there will ever exist one optimal computational strategy to generically approximate the ground state wave function. Thus, it is important to develop strategies with a wide range of features to represent the wide range of complex quantum mechanical materials that quantum computers are expected to explore.

To find the ground state of a ``target" Hamiltonian  $\mathcal{H}_\text{targ}$, an obvious strategy is to mimic nature on a quantum computer by coupling $\mathcal{H}_\text{targ}$ to a ``heat bath" Hamiltonian $\mathcal{H}_\text{bath}$ initiated in a low energy state.  For a finite quantum computer with unitary evolution, this will not produce a cold thermal state, but over a finite length of time could entangle the target system in such a sufficiently complicated way with the bath such that the von Neumann entropy of the target system, when traced over the heat bath, is peaked at the ground state of $\mathcal{H}_\text{targ}$.  However, in order for this to be useful, each eigenstate of $\mathcal{H}_\text{targ}$ must be able to efficiently transfer its energy to the heat bath. Without {\it a priori} knowledge of the eigenvectors of $\mathcal{H}_\text{targ}$ this appears in practice to require a much larger number of degrees of freedom in $\mathcal{H}_\text{bath}$ than in $\mathcal{H}_\text{targ}$. Thus, this approach is likely to scale poorly with the size of the problem and be expensive to implement.

\begin{figure}[t]
 \includegraphics[width = 0.48\textwidth]{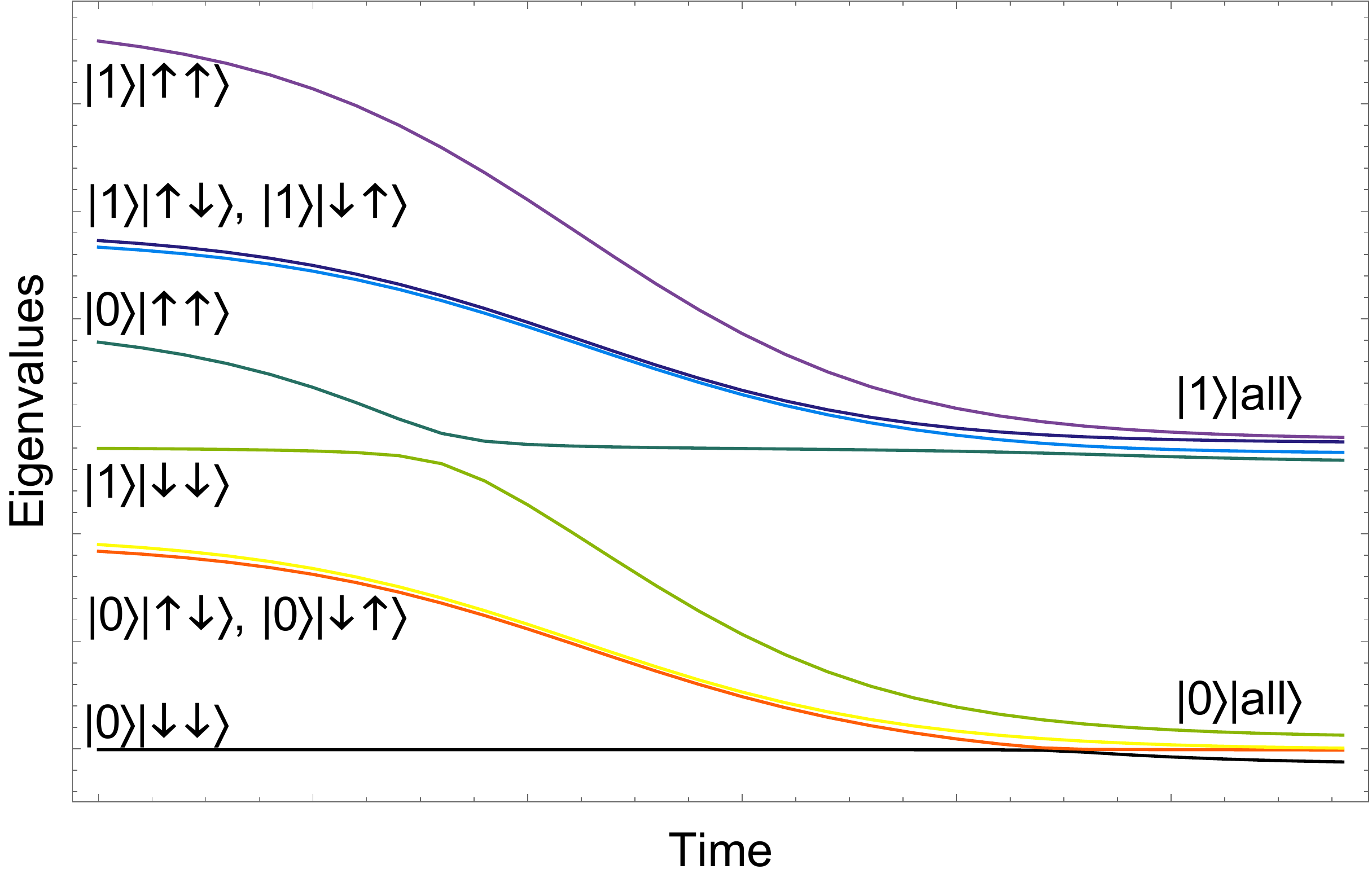}
  \caption{A simple example with $N_\text{targ}=1$, $N_\text{comb}=2$ showing how an initial state of the form $(\alpha\ket{0} + \beta\ket{1})\ket{\downarrow\downarrow}$ will adiabatically evolve into a state $\ket{0}\ket{\chi}$ after scaling the spectrum of the comb to zero energy, where $\ket{\chi}$ is some linear combination of spin configurations in the comb.
}
  \label{fig:spectrum}
\end{figure}

In this Letter, we describe a strategy that is inspired by the heat bath scenario but instead couples $\mathcal{H}_\text{targ}$ to a simple system $\mathcal{H}_\text{comb}$ which has a comparable number of degrees of freedom and a characteristic energy scale that monotonically decreases in time.
Its spectrum is, in effect, dragged through the spectrum of the target Hamiltonian $\mathcal{H}_\text{targ}$, and the avoided level crossings resonantly transfer the energy from the initial state of $\mathcal{H}_\text{targ}$ to the comb.
We refer to this algorithm as ``Spectral Combing".   A toy example with an eight-dimensional Hilbert space is illustrated in \fig{spectrum}.  The total Hamiltonian can be written as $\mathcal{H}_\text{tot}=\mathcal{H}_\text{targ}+\mathcal{H}_\text{comb}+ \mathcal{H}_\text{int}$.  Here, $\mathcal{H}_\text{targ}$ describes a 2-state system represented by a $N_\text{targ}=1$ qubit with ground state $\ket{0}$ and the excited state  $\ket{1}$.  The comb $\mathcal{H}_\text{comb}$ is 4-dimensional and realized on $N_\text{comb}=2$  qubits with approximate eigenstates  $\ket{\uparrow\uparrow}$, $\ket{\uparrow\downarrow}$, $\ket{\downarrow\uparrow}$, and $\ket{\downarrow\downarrow}$, the last being  the ground state for the comb; the  energies of these states are time dependent and monotonically decreasing. The Hamiltonian $\mathcal{H}_\text{int}$ provides a weak interaction between the target and comb. \fig{spectrum} shows how an initial state $\ket{\psi}\ket{\downarrow\downarrow}$ will evolve adiabatically into the state $\ket{0}\ket{\chi}$, where  $\ket{\psi}$ is an arbitrary state in the Hilbert space of the target Hamiltonian $\mathcal{H}_\text{targ}$, while $\ket{\chi}$ is some excited state in the Hilbert space of  $\mathcal{H}_\text{comb}$.  In this manner, the ground state of $\mathcal{H}_\text{targ}$ can be studied by measuring expectation values of operators  in the final state that act trivially on $\ket{\chi}$. While this simple model only exhibits a single avoided level crossing, larger systems will have a number that grows with the size of the Hilbert space.

The distinguishing benefits of Spectral Combing include:
(i)
One can initialize the quantum computer in a state $\ket{\psi_i}$ that is far from the ground state of $\mathcal{H}_\text{targ}$ and less costly. In fact, we anticipate that Spectral Combing will serve well as a preconditioner for initializing states for QPE;
(ii)
While the avoided  level crossings are only effective at resonantly transferring energy to the comb if the minimum energy gaps are sufficiently large, the number of such level crossings grows with the size of the Hilbert space and the effectiveness of the method does not rest on any one energy gap being large, as it does for the QAA. Thus, it is plausible, with supporting evidence from the explicit  simulation of small systems discussed below,  that the computational cost of the method will scale as a reasonable  power of the number of qubits $N$  needed to represent ${\mathcal H}$;
(iii)
The output of one Spectral Combing can become the input (trial wavefunction) of a subsequent combing, so that the result can be iteratively improved (at the expense of requiring a longer coherence time).

In the next sections, we describe Spectral Combing in some detail, and give the results of simulations for small systems.

\section{Method}

The comb we construct here is  a bosonic spin system governed by the following Hamiltonian  on $N_c$ qubits,
\begin{equation}
\mathcal{H}_\text{comb}(t) = \sum_{i = 1}^{N_c}\nu(t)
 \sigma_i^+ \sigma_i^-
+ \kappa\phi_i\sum_{cyc}
\left(\sigma_i^+\sigma_{i+1}^-\sigma_{i+2}^- + h.c. \right)\ .
\nonumber
\end{equation}
The function $\nu(t)$  monotonically decreases with time, and when ${\cal H}_\text{comb}$ and  ${\cal H}_\text{targ}$ are weakly coupled, gives rise to a large number of avoided  level crossings (see \fig{spectrum}). In this preliminary study, we use the simple linear form
\beq
\nu(t) = \nu_0(1-t/t_f) \ ,\qquad \nu_0>0\ ,\quad 0\le t\le t_f\ .
\nonumber
\eeq
The interaction term proportional to coupling constant $\kappa$, with the $\phi_i$ being real random numbers,  is included to break symmetries with the aim of making the system ergodic and to minimize ÒscarringÓ, allowing the comb to couple efficiently to a generic target system.
This interaction can be thought of as a $1\leftrightarrow 2$ particle scattering interaction, allowing energy to disperse within the comb subsystem. The summation runs over the cyclic permutations of the nearest neighboring spins only. 
For small $\kappa$, the approximate ground state of the comb is $\ket{\downarrow\ldots\downarrow}$.

In order to entangle the comb and the target system, represented by $N_t$ qubits, the Hamiltonian
\begin{equation}
\mathcal{H}_\text{int} = g\left[\mathbf{A} \otimes \sum_{ i }^{N_c} (\sigma_i^+ + \sigma_{i+1}^++\sigma_i^+\sigma_{i+1}^+) \right]+ h.c.
\label{eq:hint}
\end{equation}
is used, which allows energy to flow between the two systems. The operator $\mathbf{A}$  is designed to couple a significant number of states in the target to
 those of the comb, and must not commute with $\mathcal H_\text{targ}$. In the next section, where we consider the Ising model, we have  found satisfactory results with good scaling properties by taking $\mathbf{A}$ to equal either a sum of one-body operators or a random matrix.

Starting from an initial  state $\ket{\psi}\ket{\downarrow\downarrow\cdots}$, where $\ket{\psi}$ is an arbitrary state in the space of ${\mathcal H}_\text{targ}$ while  $\ket{\downarrow\downarrow\cdots}$ is the approximate ground state of ${\mathcal H}_\text{comb}$ (exact as $\kappa\to 0$), our algorithm proceeds  as follows:
\begin{enumerate}
 \item propagate the state from $t=0$ to $t=t_f$ using the total Hamiltonian $\mathcal{H} = \mathcal{H}_\text{targ} + \mathcal{H}_\text{comb} + \mathcal{H}_\text{int}$ and time step $\delta t$;
 \item if the predetermined maximum number of iterations is reached then exit; otherwise perform a measurement of the $z$-projection of the spins in the comb in order to collapse the wave function to a pure state in the comb;
 \item return the spins in the comb to their ground state, rescale $ g\to\eta g$ and  $\nu_0\to\eta \nu_0$ with $\eta<1$, and repeat.
\end{enumerate}

 Six  parameters $\left\{\nu_0,t_f,\kappa,g,\delta t,\eta\right\}$ in this algorithm may be chosen to optimize its performance. In this preliminary study,  we chose these parameters to approximately minimize the energy of the final state for a given number of comb iterations, each of which was a fixed number of Trotter steps $N_T=t_f/\delta_t$.
In general, this procedure can be performed using an automatic optimization algorithm that can run on a classical computer.

\section{\label{sec:Results}Results}

To efficiently study the properties of Spectral Combing, we computed the time evolution of the wave function both by emulation and by simulation of the quantum circuit given in the supplemental material. Emulation allows for better computational performances in narrowing down optimal variational parameters   \cite{HPemulation}  but, in the long run, it may be more informative to closely simulate the circuits used on a quantum computer (i.e. being consistent with the chosen approximation of the time-evolution operator) in order to maximize the overall efficiency.

\begin{figure}
 \ig[width = 0.45\textwidth]{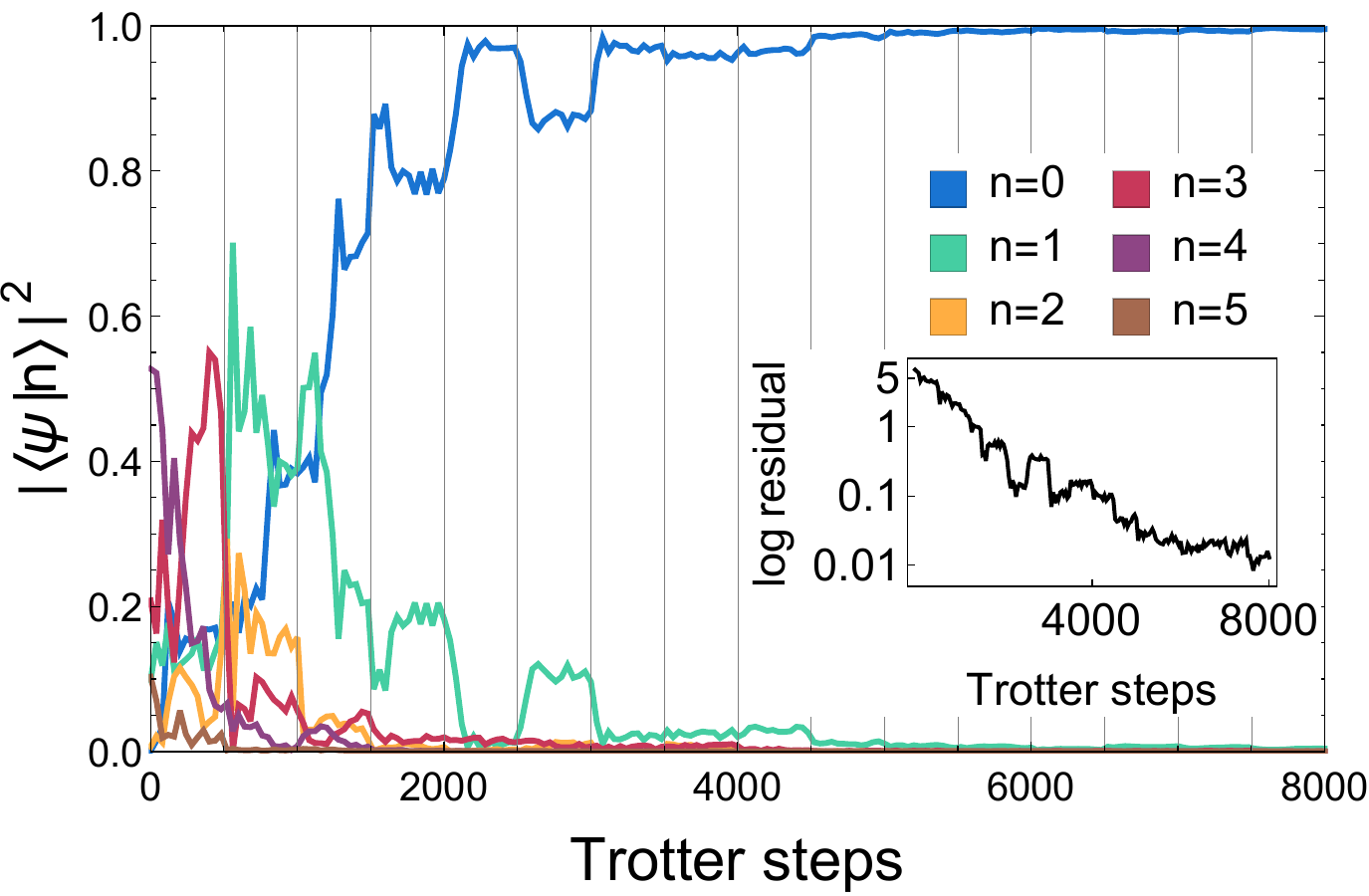}
 \caption{The emulated evolution of the target's overlap with the six lowest eigenstates of the 3-spin Ising model $|\psi_n\rangle$ coupled to a 3-qubit comb, starting from a  random initial state.  Vertical lines represent iteration every $500$ Trotter steps. The inset shows the decreasing residual between the target energy and that of the true ground state.}
 \label{fig:Ising3x3}
 \end{figure}

In order to assess the performance of the Spectral Combing procedure, we first tested it for the 1D transverse field Ising model described by the Hamiltonian
\begin{equation}
\mathcal{H}^I_\text{targ}=-h\sum_{i = 1}^{N_t}\sigma_i^x - \sum_{i=1}^{N_t} \sigma_i^z \sigma_{i+1}^z\ .
\label{eq:Ising1d}
\end{equation}
For $h=0$ this theory has two degenerate ground states, the states with all spins aligned either up or down; for small nonzero $h$ these two mix and the ground state is split from the first excited state by a gap $\Delta = O(h^{N_t})$.   \fig{Ising3x3} shows an example   for $h=2.0$ of the time evolution of the overlap of the target wave function with the lowest six eigenstates of ${\mathcal H_\text{targ}}$,  where each of the combings takes 500 time steps and is followed by a measurement of the comb spins (step 2 above).  One can see that, even though the randomly chosen initial state has only a 0.1\% overlap with the true ground state, it evolves eventually to have a 99\% overlap (blue line), with continued improvements with each successive iteration.

\begin{figure}
 \ig[width = 0.45\textwidth]{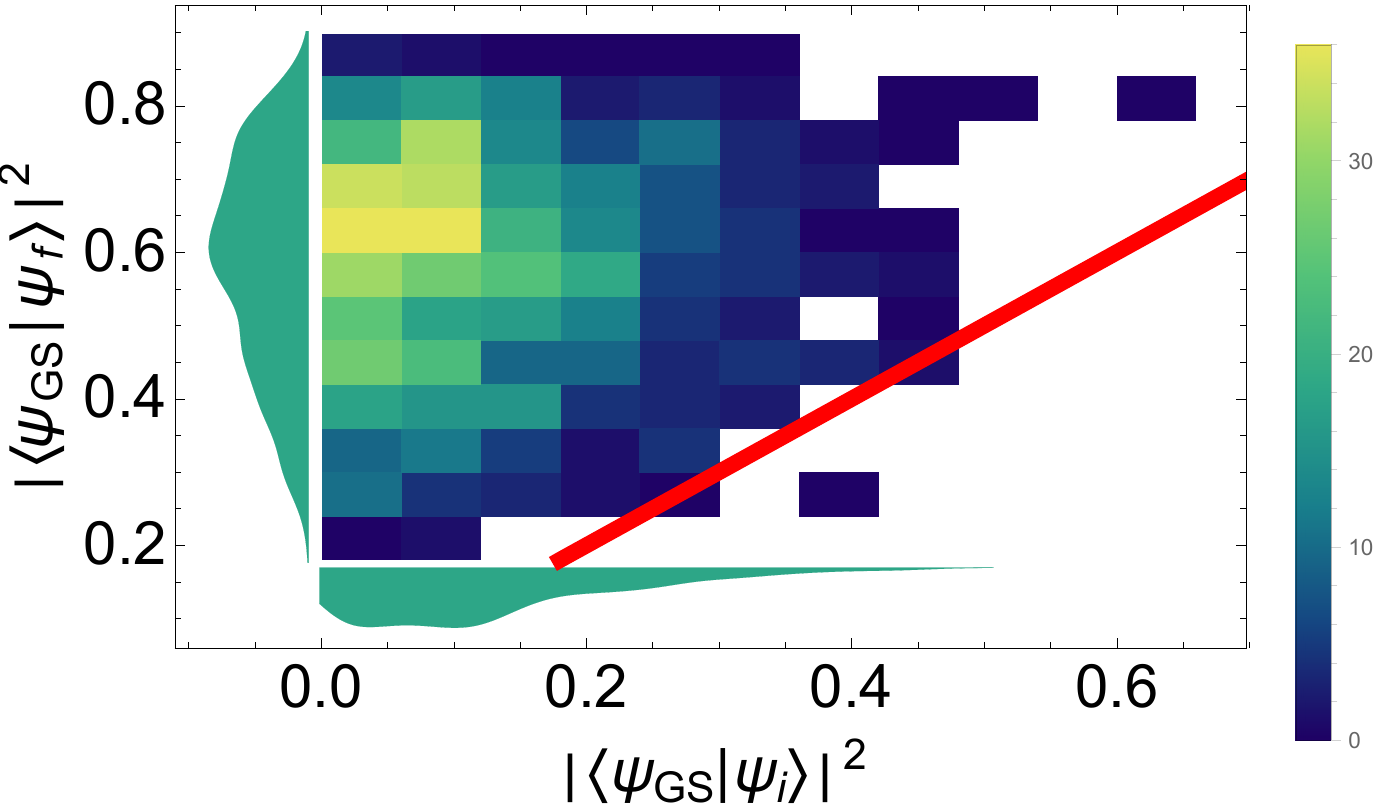}
 \caption{Limited to 2000 Trotter steps distributed over 2 combs, the improvement of the ground state overlap for the 3-spin Ising model with $h = 0.5$ is shown for a sample of 800 possible initial states and comb measurements. The diagonal red line represents no improvement.}
 \label{fig:success}
 \end{figure}

Although it is theoretically satisfying that Spectral Combing with unlimited resources achieves better than 1\% precision of both the ground state energy and overlap with the true ground state, it is more efficient to use this method to obtain a less precise approximation to the ground state, and then subsequently refine it by QPE.  In this strategy, the Hilbert space is enlarged for Spectral Combing while knowledge of the ground state is minimal, and then reduced for QPE to efficiently improve precision.  It is this consideration of resources that  leads us to define  the benchmark of success for the application of Spectral Combing to be $\left\vert\braket{\psi_\text{gs}}{\psi_f}\right\vert^2\approx 0.5$.  \fig{success} shows the generic ability of this method to achieve such a definition of success with a moderate number of Trotter steps for the Ising model Eq.~\eqref{eq:Ising1d}, even with initial wave functions nearly orthogonal to the ground state. For all cases considered with different values of the external field $h$ spanning both sides of the phase transition at $h$ = 1 and  $N_t = 3,4$, and $5$ qubits, we found it sufficient to use a carefully optimized system of only $3$ qubits in the comb with $\mathcal{O}\left(10^3\right)$ Trotter steps growing approximately linearly with $N_t$. These results have been obtained employing a matrix $\mathbf{A}$ from Eq.~\eqref{eq:hint} with random real entries where $\mathcal{H}_\text{targ}$ is nonzero.

An application for Spectral Combing could be for physical systems
where a good trial wave function for the ground state is not known, and where it is not known how to employ Quantum Adiabatic Algorithm without encountering small energy gaps along the adiabatic path.  In order to demonstrate the advantage of Spectral Cooling in this case with a small and tractable number of qubits, we consider the Ising model of Eq.~\eqref{eq:Ising1d} with an additional magnetic field $B$ applied in the $z$ direction \footnote{We thank M. Troyer for suggesting his model for comparison.},
\begin{equation}
\mathcal{H}^{B}_\text{targ}=-h\sum_{i = 1}^{N_t}\sigma_i^x - \sum_{i = 1}^{N_t}\sigma_i^z\sigma_{i+1}^z + B \sum_{i = 1}^{N_t}\sigma_i^z\ .
\label{eq:Ising1dBfield}
\end{equation}
For large   positive/negative $ B$, the ground state is approximately all spins aligned down/up.  There is an energy barrier between these states characterized by the small gap $\Delta=O(h^{N_t})$ at $B=0$ and as $N_t\to\infty$ the model exhibits a first order phase transition as $B$ is varied through zero. When taking $B$ as the adiabatic parameter for the QAA along the path from $B=+1$ to $B=-1$, one can expect the required number of Trotter steps  to scale as $1/\Delta^2$.   We compare the cost of such a QAA computation as a function of $\Delta$ (with linear interpolation in $B$) with that for a Spectral Combing computation where we fix $B=-1$, but choose as our initial wave function the exact ground state of  $\mathcal{H}^{B}_\text{targ}$ for $B=+1$, a wave function that corresponds to the metastable state on the wrong side of the large energy barrier.  We are not claiming that this is the optimal application of the QAA to such a problem, nor that this is the best initial wave function one could use for Spectral Combing, but wish to compare the performance of these two algorithms under controlled non-optimal conditions.

\begin{figure}[t]
 \ig[width = 0.45\textwidth]{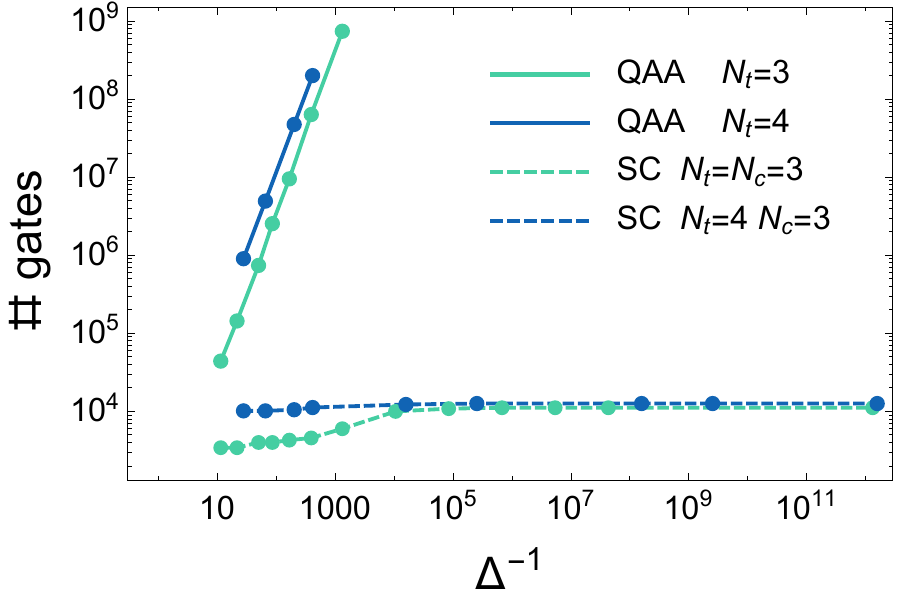}
 \caption{Results for the Ising model with external magnetic field for both the Quantum Adiabatic Algorithm (QAA) and Spectral Combing using $N_c=3$ spins in the comb. The gate estimates per Trotter step are detailed in the supplemental material and calculated to be 21(28) for the QAA, and 283(347) for the spectral comb simulation employing $N_t=3$($N_t=4$).}
\label{fig:IsingBfield}
\end{figure}

  The results of this comparison for the two cases $N_t = 3, 4$ is shown in \fig{IsingBfield}, plotting  versus $1/\Delta$ the cost in number of gates required to achieve  $\gtrsim 50\%$ overlap with the true ground state at $B=-1$. For both methods, we estimate the number of quantum gates required to perform the evolution using the complete implementation of the algorithm presented in the supplemental material.
 For the QAA the cost per Trotter time-step is given simply by the number of gates needed to implement the propagator for $\mathcal{H}^{B}_\text{targ}$ while for the SC method the gate cost must also account for the propagators of $\mathcal{H}_\text{comb}$ and $\mathcal{H}_\text{int}$. Compared to the results in \fig{success} the cost in Trotter steps was greatly reduced down to $\mathcal{O}(50)$ thanks to the use of a different coupling matrix $\mathbf{A}$ written as a sum of Pauli $\sigma^x$ matrices and by a more careful optimization of the comb parameters, made possible by specializing to a particular initial vector.

As shown in  \fig{IsingBfield}, the gate cost for the QAA method scales proportional to $\Delta^{-2}$ as expected, and becomes  prohibitive for very small $\Delta$. In contrast, Spectral Combing exhibits superior scaling with the inverse gap, despite the doubling of qubits required to create the comb and having started from a metastable state separated from the ground state by a large barrier, becoming surprisingly independent of $\Delta$ at small values.  In all cases shown in \fig{IsingBfield} the state preparation for Spectral Combing has been achieved with a single comb, thus avoiding the cost and intrinsic randomness of the measurement process.

\section{\label{sec:Discussion}Discussion}

Spectral Combing is presented as an algorithm
for identifying  the approximate ground state  of a strongly correlated quantum mechanical system, that can in turn be profitably used as the initial state for a Quantum Phase Estimation procedure. This is a quintessentially quantum algorithm that is not suited for a classical computer, as the first step to finding the eigenvector with minimum eigenvalue for a very large matrix is to vastly increase the dimension of that matrix,  requiring an approximate doubling of the number of qubits needed to represent the Hamiltonian.  An explicit construction of the quantum circuit required for Spectral Combing is given in  Supplemental Material.
Further work needs to be done to better determine the scaling properties of this algorithm for larger system sizes, as has been performed for other algorithms (cf., Ref.~\cite{Wecker2014Gatecount}).

The properties of the excitations of a quantum many-body system are also of interest.  For example one may wish to understand the properties of nuclei directly from the theory of quarks and gluons \cite{Beane:2012vq}, or the cross section for scattering pions.  Spectral Combing could be useful for cases in which an excitation is the lightest state with a given set of quantum numbers;   a comb must be designed that respects the symmetries in question, as does its interaction with the target system, and then  the initial wave function is chosen to have the desired quantum numbers (e.g., linear or angular momentum, baryon or fermion number, electric charge, or isospin). The algorithm is then expected to find the minimum energy state carrying those quantum numbers.

A virtue of the Spectral Combing algorithm is that it relies on the spectral complexity of the comb designed to  handle diverse Hamiltonians, rather than having to engineer a particular solution for each Hamiltonian. We are optimistic that this work could contribute to the development of a practical way to access the properties of the ground state and low-lying excitations of arbitrarily complex quantum many-body systems, but further study is required.

\section{Acknowledgements}
We greatly indebted to members of the QuArC group at Microsoft Research for very useful interactions and comments on this manuscript, particularly M. Troyer and N. Wiebe.  We also thank M. Savage, K. Roche, I. Abdurrahman, and other members of the INTUW Quantum Computing Group for numerous discussions.
DK, NK and AR were supported in part by DOE Grant No. DE-FG02-00ER41132; DK was supported in part by the Thomas L. and Margo G. Wyckoff Endowed Faculty Fellowship; NK was supported in part by the Seattle Chapter of the Achievement Rewards for College Scientists (ARCS) foundation.

\bibliographystyle{apsrev4-1}
\bibliography{ResCbib} 

\appendix
\section{\label{sec:sup} SUPPLEMENTAL MATERIAL}
\section{\label{sec:Circuit} Quantum Circuit Implementation}
\par The simulation of physical quantum systems using quantum computers has been  long-awaited  \cite{Feynman}.  Inspired by recent progress in this area represented in part by Ref. \cite{Georgescu,Barends2015,Buluta108,Lanyon57,Lloyd1073,Martinez:2016yna}, we provide a full gate implementation of the time evolution operator needed for Spectral Combing with a target Hamiltonian defined by a 1D Ising model for a system of $N_t=3$ spins.
\par The gate set used is standard and has been defined in Ref. \cite{nielsen2010quantum}.  It includes the following one-qubit gates:
\begin{itemize}
 \item Hadamard
 \begin{equation*}
  \Qcircuit @C=1em @R=.7em {
& \gate{H} & \qw
}
=\frac{1}{\sqrt{2}}\begin{pmatrix}
   1 & 1 \\
   1 & -1
 \end{pmatrix}\end{equation*}
 \item z-axis rotation of angle $\theta$
 \begin{equation*}
  \Qcircuit @C=1em @R=.7em {
& \gate{e^{-i \theta \sigma_z}} & \qw
} = \begin{pmatrix}
  e^{-i \theta} & 0 \\
  0 & e^{i\theta}
\end{pmatrix}
 \end{equation*}
 \item phase gate
 \begin{equation*}
  \Qcircuit @C=1em @R=.7em {
& \gate{S} & \qw
} = \begin{pmatrix}
  1 & 0 \\
  0 & i
\end{pmatrix}
 \end{equation*}
\end{itemize}
and the two-qubit gates:
\begin{itemize}
 \item $\textit{SWAP}$
  \begin{equation*}
  \begin{gathered}
  \Qcircuit @C=1em @R=1.1em {
& \qswap & \qw \\
 & \qswap \qwx & \qw
}\end{gathered} = \begin{pmatrix}
  1 & 0 & 0 & 0 \\
  0 & 0 & 1 & 0 \\
  0 & 1 & 0 & 0 \\
  0 & 0 & 0 & 1
\end{pmatrix}
 \end{equation*}
 \item $\textit{CNOT}$
 \begin{equation*}
 \begin{gathered}
\Qcircuit @C=1em @R=.7em {
& \ctrl{1}  & \qw \\
& \targ &  \qw
}\end{gathered} = \begin{pmatrix}
  1 & 0 & 0 & 0 \\
  0 & 1 & 0 & 0 \\
  0 & 0 & 0 & 1 \\
  0 & 0 & 1 & 0
\end{pmatrix}
 \end{equation*}
\end{itemize}
\par Though there are known alternatives with attractive scaling properties \cite{BerryChilds_TaylorSeries,WiebeChilds}, first-order Trotterization\cite{Trotter1959} is used here to evolve the state in time. The complete circuit is thus a representation of the propagator for a single Trotter step at time $t$.

\subsection{ Target Propagator }
We start with the implementation of the Trotterized propagator associated with $\mathcal{H}_\text{targ}$ (see main text Eq.~(2)):
\begin{equation}
\mathcal{H}^I_\text{targ}=-h\sum_{i = 1}^{N_t}\sigma_i^x - \sum_{i=1}^{N_t} \sigma_i^z \sigma_{i+1}^z\;.
\end{equation}
If the target Hamiltonian of Eq.~(3) of the main text was chosen where a $B$-field is added in the $z$-direction, only an additional $N_t$ gates would be needed to propagate the target space.
\par Quantum circuits act on qubits from left to right and we thus implement operators forming the Trotterized propagator in Eq.~\eqref{eq:targprop} from right to left.
\begin{widetext}
\begin{align}
e^{-i \mathcal{H}_\text{targ}\delta t} &\rightarrow e^{i \delta t \sigma_1^z \sigma_3^z} e^{i \delta t \sigma_2^z \sigma_3^z} e^{i \delta t \sigma_1^z \sigma_2^z} e^{i h \delta t \sigma_1^x} e^{i h \delta t \sigma_2^x} e^{i h \delta t \sigma_3^x} \label{eq:targprop}\\ &= \qquad
\begin{gathered}
\scalebox{0.7}{
\Qcircuit @C=.7em @R=0.1cm @! {
\lstick{\ket{\psi}_1} & \gate{H} & \gate{e^{ih \delta t \sigma_z}} & \gate{H} & \ctrl{1} & \qw & \ctrl{1} & \qw  & \qw & \qw & \ctrl{2} & \qw & \ctrl{2} & \qw \\
\lstick{\ket{\psi}_2} & \gate{H} & \gate{e^{ih\delta t\sigma_z}} & \gate{H} & \targ & \gate{e^{i\delta t\sigma_z}} & \targ & \ctrl{1} & \qw & \ctrl{1} & \qw  & \qw & \qw & \qw \\
\lstick{\ket{\psi}_3} & \gate{H} & \gate{e^{ih\delta t\sigma_z}} & \gate{H} & \qw & \qw  & \qw & \targ & \gate{e^{i\delta t\sigma_z}} & \targ & \targ & \gate{e^{i\delta t\sigma_z}} & \targ & \qw
}
}
\label{eq:targcirc}
\end{gathered}
\end{align}
\end{widetext}
\par The 1-body terms may be simulated in parallel (first 3 gates acting on each qubit) as they are commuting operators acting in separate spaces. In order to evolve each qubit under the effect of the transverse field, the following standard steps are implemented:
\begin{itemize}
 \item perform a rotation from the $z$- to the $x$-basis using a Hadamard gate
 \item apply a $z$-rotation with angle $\theta=-h \delta t$
 \item rotate the axis back to the $z$-basis with a final Hadamard gate
\end{itemize}
\par The remaining 2-body interaction terms $\sigma_i^z\sigma_j^z$ do not require any basis change. The next set of three gates represents the evolution of this 2-body interaction for the pair $(1,2)$. The standard 1-qubit, $z$-axis rotation with $\theta = -\delta t$ is surrounded with $\textit{CNOT}$ gates to entangle the two qubits. The role of these $\textit{CNOT}$s here can be summarized as compactly managing the multiple minus signs that appear when taking the tensor product of $\sigma^z$s in the exponent.  Thinking of the $\textit{CNOT}$ gates as controlled-$\sigma_x$ gates and recalling the anti-commutation relation of Pauli matrices giving $\sigma_x\sigma_z\sigma_x = -\sigma_z$ is sufficient to rationalize the function of this 3-gate combination.
\par These two elaborations on the simple $z$-axis rotation gate---basis changes and multi-qubit entanglement---will be sufficient to understand the circuitry for the remaining contributions to the propagator in its entirety.

\subsection{ Comb Propagator }
\par The propagator for a 3-spin comb will be governed by the Hamiltonian (see Eq.~(1) of the main text)
\begin{equation}
\begin{split}
&\mathcal{H}_\text{comb}(t) = \sum_{i}^{3}\nu(t)
 \sigma_i^+ \sigma_i^- \\
&\hspace{1cm} + \kappa \phi \sum_{cyc}
\left(\sigma_4^+\sigma_5^-\sigma_6^- + h.c. \right)
\end{split}
\label{eq:Hcomb}
\end{equation}
and may be expressed in quantum gates as:
\begin{widetext}
\begin{align}
e^{-i H_{comb}(t) \delta t} &\rightarrow  e^{-i \frac{\kappa \phi \delta t}{4} \sigma_4^y\sigma_5^y \sigma_6^x} e^{-i \frac{\kappa \phi\delta t}{4} \sigma_4^y\sigma_5^x \sigma_6^y} e^{-i \frac{\kappa \phi\delta t}{4} \sigma_4^x\sigma_5^y \sigma_6^y} e^{-i \frac{3\kappa \phi\delta t}{4} \sigma_4^x\sigma_5^x \sigma_6^x} \\& \hspace{6cm} \times e^{-i\nu(t) \frac{1}{2}\sigma_6^z\delta t} e^{-i\nu(t) \frac{1}{2}\sigma_5^z\delta t} e^{-i\nu(t) \frac{1}{2}\sigma_4^z\delta t} \nonumber\\
&\rightarrow \qquad
\begin{gathered}
\scalebox{0.65}{
\Qcircuit @C=-1.2em @R=-1.cm @! {
\lstick{\ket{0}_4} & \gate{e^{-i\nu(t) \delta t\frac{1}{2}\sigma_z}}  \\
\lstick{\ket{0}_5} & \gate{e^{-i\nu(t) \delta t\frac{1}{2}\sigma_z}} \\
\lstick{\ket{0}_6} & \gate{e^{-i\nu(t) \delta t\frac{1}{2}\sigma_z}}
}
}
\hspace{-0.06cm}
\scalebox{0.65}{
\Qcircuit @C=-1.9em @R=-0.8cm @! {
  & \gate{H} & \ctrl{1} & \qw& \qw & \qw & \ctrl{1} & \qw & \qw  & \ctrl{1} & \qw & \qw & \qw & \ctrl{1} &  \gate{H}\\
  &\gate{H} & \targ & \ctrl{1}& \qw & \ctrl{1} & \targ & \gate{H} & \gate{S^\dagger H} & \targ & \ctrl{1} & \qw & \ctrl{1} & \targ & \gate{HS}\\
   &\gate{H} & \qw  & \targ& \gate{e^{-i\frac{3 \kappa \phi\delta t}{4}\sigma_z}} & \targ & \qw & \gate{H} & \gate{S^\dagger H} & \qw & \targ & \gate{e^{-i\frac{\kappa \phi \delta t}{4}\sigma_z}} & \targ & \qw & \qw
}}
\\\vspace{0.5cm} \\
\scalebox{0.65}{
\Qcircuit @C=-1.1em @R=-0.8cm @! {
\lstick{\ket{\cdots}_4} & \gate{S^\dagger H} & \ctrl{1} & \qw & \qw & \qw & \ctrl{1} & \qw & \qw & \ctrl{1} & \qw & \qw & \qw & \ctrl{1} & \gate{HS}\\
\lstick{\ket{\cdots}_5} & \gate{H} & \targ & \ctrl{1} & \qw & \ctrl{1} & \targ & \gate{H} & \gate{S^\dagger H} &  \targ & \ctrl{1} & \qw & \ctrl{1} & \targ & \gate{HS}\\
\lstick{\ket{\cdots}_6} & \qw  & \qw & \targ & \gate{e^{-i\frac{\kappa \phi\delta t}{4}\sigma_z}} & \targ & \qw & \gate{HS} & \gate{H} & \qw & \targ & \gate{e^{-i\frac{\kappa \phi\delta t}{4}\sigma_z}} & \targ & \qw & \gate{H}
}}
\end{gathered}
\end{align}
\end{widetext}
\par In order to propagate the state of the comb with the $3$-spin interaction, the phase gate $S$ is used (in combination with the Hadamard) for transforming from the $z$- to the $y$-basis.
\par It is interesting to note that the inclusion of cyclic permutations of this interaction does not add gates to the cost of the circuit but only modifies the values of rotation angles controlling existing gates.  Introducing different coefficients for each permutation is also a choice that introduces no new gates into the circuit. While this would make the comb spectrum more dense with additional broken degeneracies, the current simulations have not needed the additional level-repulsions and thus a single coefficient, $\phi$, was used for simplicity.
\par Because the target and comb propagators may be written as tensor product operators in the target-comb space, they may be simulated in parallel (assuming capable hardware).  Such parallelization becomes advantageous as it diminishes the amount of time the quantum state must be isolated from its environment.  With the current choice of target and construction of the comb, the target gate count scales as $N_t$ while the comb scales as $N_c$ making parallelization an effective removal of a sub-leading $\mathcal{O}(N)$ contribution to simulation time.

\subsection{ Interaction Propagator }
In the current implementation, terms in the comb-target interaction Hamiltonian have been limited to nearest-neighbor, two-body interactions in the comb coupled to one-body terms in the target (see Eq.~(2) of the main text) which may be written for this $N_t = N_c = 3$ example as:
\begin{widetext}
\begin{equation}
\begin{split}
\mathcal{H}_\text{int} &= g\left[\mathbf{A} \otimes \sum_{ i }^{3} (1 + \sigma_i^+)(1 + \sigma_{i+1}^+)-1 \right]+ h.c.\\
&=g \left[\left(-h\sum_{i= 1}^{3}\sigma_i^x\right)\otimes 2\sigma_4^x + 2\sigma_5^x + 2\sigma_6^x  +\frac{1}{2}\left(\sigma_4^x\sigma_5^x - \sigma_4^y\sigma_5^y + \sigma_5^x\sigma_6^x - \sigma_5^y\sigma_6^y + \sigma_4^x\sigma_6^x - \sigma_4^y\sigma_6^y\right)\right]
\end{split}
\end{equation}

In this case, the number of gates needed to implement each Trotter step of the interaction propagator scales asymptotically as $N_tN_c$. The circuit implementation of the interaction propagator is:
\begin{equation}
e^{-i \mathcal{H}_\text{int}\delta t} \rightarrow
\qquad
\begin{gathered}
\scalebox{0.8}{
\Qcircuit @C=0.2em @R=0.0cm @! {
\lstick{\ket{\cdots}_t} & \qw &  \multigate{3}{A_1}  & \qw & \qw & \qw & \multigate{3}{A_1} & \qw & \qw & \qw & \qw & \multigate{3}{A_1} & \qw & \qw & \multigate{4}{A_2}  & \qw & \qw \\
\lstick{\ket{\cdots}_t} & \qw &  \ghost{A_1}  & \qw & \qw & \qw & \ghost{A_1} & \qw & \qw & \qw & \qw & \ghost{A_1} & \qw & \qw & \ghost{A_2} & \qw & \qw\\
\lstick{\ket{\cdots}_t} &\qw  & \ghost{A_1}  & \qw & \qw & \qw & \ghost{A_1} & \qw & \qw & \qw& \qw & \ghost{A_1} & \qw & \qw & \ghost{A_2} & \qw & \qw  \\
\lstick{\ket{\cdots}_c} & \gate{H} & \ghost{A_1} & \gate{H} & \qswap & \gate{H} & \ghost{A_1} & \gate{H} & \qw & \qswap & \gate{H} & \ghost{A_1}  & \qswap & \gate{H} &  \ghost{A_2} & \gate{H} & \qw \\
\lstick{\ket{\cdots}_c} & \qw & \qw &\qw & \qswap \qwx & \qw & \qw & \qw & \qswap & \qswap \qwx &  \qw & \qw &  \qswap \qwx & \qw & \ghost{A_2^+} & \gate{H} & \qw\\
\lstick{\ket{\cdots}_c}  & \qw &\qw & \qw & \qw & \qw & \qw & \qw  & \qswap \qwx & \qw  & \qw & \qw & \qw & \qw & \qw & \qw & \qw
}
}
\\ \\
\scalebox{0.7}{
\Qcircuit @C=0.1em @R=-0.1cm @! {
\lstick{\ket{\cdots}_t} & \qw  & \multigate{4}{A_2^-}  & \qw & \qw & \qw & \qw & \multigate{4}{A_2^+} & \qw & \qw & \multigate{4}{A_2^-} & \qw &\qw & \qw & \multigate{4}{A_2^+} & \qw & \qw & \multigate{4}{A_2^-} &\qw & \qw\\
\lstick{\ket{\cdots}_t} & \qw  & \ghost{A_2} & \qw & \qw & \qw & \qw & \ghost{A_2} & \qw & \qw & \ghost{A_2} & \qw &\qw & \qw & \ghost{A_2} & \qw & \qw & \ghost{A_2}&\qw & \qw\\
\lstick{\ket{\cdots}_t} & \qw  & \ghost{A_2}  & \qw & \qw & \qw & \qw & \ghost{A_2} & \qw  &\qw & \ghost{A_2} & \qw&\qw & \qw &  \ghost{A_2} & \qw & \qw & \ghost{A_2} & \qw & \qw \\
\lstick{\ket{\cdots}_c}  & \gate{S^\dagger H} &\ghost{A_2}& \gate{H S} & \qw & \qswap & \gate{H} & \ghost{A_2} & \gate{H} & \gate{ S^\dagger H} & \ghost{A_2} & \gate{H S} & \qw &\gate{H} & \ghost{A_2} & \gate{H} & \gate{S^\dagger H} & \ghost{A_2} & \gate{H S} & \qw\\
\lstick{\ket{\cdots}_c} &  \gate{S^\dagger H} & \ghost{A_2} & \gate{H S} &\qswap & \qswap \qwx & \gate{H} & \ghost{A_2} & \gate{H} & \gate{S^\dagger H} &  \ghost{A_2} & \gate{HS} & \qswap &  \gate{H} & \ghost{A_2} & \gate{H} & \gate{S^\dagger H} &  \ghost{A_2} & \gate{HS} & \qswap\\
\lstick{\ket{\cdots}_c} &\qw  & \qw & \qw & \qswap \qwx & \qw & \qw &\qw & \qw & \qw  & \qw & \qw  & \qswap \qwx & \qw &\qw  & \qw  & \qw & \qw & \qw  & \qswap \qwx
}
}
\end{gathered}
\end{equation}
with the following passages to implement the target-space Hamiltonian in which the one-body part of the target Hamiltonian is simulated as:
\begin{equation}
\begin{gathered}
\scalebox{0.7}{
\Qcircuit @C=0.8em @R=-0.1cm @! {
& \multigate{3}{A_1} & \qw \\
 & \ghost{A_1} & \qw \\
 & \ghost{A_1} & \qw \\
 & \ghost{A_1} & \qw
}}
\end{gathered} =
\qquad
\begin{gathered}
\scalebox{0.7}{
\Qcircuit @C=-0.8em @R=-0.5cm @! {
\lstick{\ket{\cdots}_t} & \gate{H} & \ctrl{3} &  \qw& \ctrl{3} & \qw & \qw & \qw & \qw & \qw & \qw & \gate{H} & \qw \\
\lstick{\ket{\cdots}_t} & \gate{H} & \qw &  \qw & \qw &\ctrl{2} &  \qw & \ctrl{2} & \qw & \qw &  \qw & \gate{H} & \qw \\
\lstick{\ket{\cdots}_t} & \gate{H}  & \qw & \qw & \qw & \qw & \qw & \qw & \ctrl{1} & \qw  & \ctrl{1} & \gate{H}& \qw \\
\lstick{\ket{\cdots}_c} & \qw  & \targ & \gate{e^{2ihg \delta t\sigma_z}} & \targ & \targ & \gate{e^{2ihg \delta t\sigma_z}} & \targ & \targ & \gate{e^{2ihg \delta t\sigma_z}} & \targ & \qw & \qw
}}
\end{gathered}
\end{equation}
\begin{equation}
\begin{gathered}
\scalebox{0.7}{
\Qcircuit @C=0.8em @R=-0.1cm @! {
& \multigate{4}{A_2^\pm} & \qw \\
 & \ghost{A_2^\pm} & \qw \\
 & \ghost{A_2^\pm} & \qw \\
 & \ghost{A_2^\pm} & \qw \\
 & \ghost{A_2^\pm} & \qw
}}
\end{gathered} =
\qquad
\begin{gathered}
\scalebox{0.7}{
\Qcircuit @C=-1.7em @R=-1.0cm @! {
\lstick{\ket{\cdots}_t} & \gate{H} & \ctrl{3} & \qw & \qw & \qw& \ctrl{3} & \qw & \qw & \qw & \qw & \qw & \qw & \qw & \qw & \qw & \qw & \gate{H} & \qw \\
\lstick{\ket{\cdots}_t} & \gate{H} & \qw & \qw & \qw & \qw & \qw &\ctrl{2} & \qw &\qw & \qw & \ctrl{2} & \qw & \qw & \qw & \qw & \qw & \gate{H} & \qw \\
\lstick{\ket{\cdots}_t} & \gate{H} & \qw & \qw & \qw & \qw & \qw & \qw & \qw & \qw & \qw & \qw & \ctrl{1} & \qw & \qw & \qw & \ctrl{1} & \gate{H}& \qw \\
\lstick{\ket{\cdots}_c} & \qw  & \targ & \ctrl{1} & \qw & \ctrl{1} & \targ & \targ & \ctrl{1} & \qw & \ctrl{1} & \targ & \targ & \ctrl{1} & \qw & \ctrl{1} & \targ & \qw & \qw\\
\lstick{\ket{\cdots}_c} & \qw & \qw & \targ & \gate{e^{\pm\frac{i}{2}hg \delta t\sigma_z}} & \targ & \qw & \qw & \targ & \gate{e^{\pm\frac{i}{2}hg \delta t\sigma_z}} & \targ & \qw & \qw & \targ & \gate{e^{\pm\frac{i}{2}hg \delta t\sigma_z}} & \targ & \qw & \qw & \qw
}}
\end{gathered}
\end{equation}
\end{widetext}

\section{Scaling}
With the above gate construction and intended extension to larger systems, the simulation of a single Trotter step using the Spectral Comb has the following resource requirements:
\begin{widetext}
\begin{equation}
  \text{gates}(N_t,N_c) = \overbrace{6N_t}^{\text{target}} + \overbrace{N_c  + 46 \begin{cases}1 & \text{if } N_c = 3 \\ N_c & \text{if } N_c >3\end{cases}}^\text{comb} +
  \overbrace{N_c\left(\left[5N_t\right]_{A_1} + 2 \left[7N_t\right]_{A_2}+ 14\right) }^\text{interaction} \quad \stackrel{\mathlarger\sim}{\scalebox{0.8}{\tiny\text{large N}}} \quad N_cN_t
\end{equation}
\begin{equation}
  \text{rotations}(N_t,N_c) = \overbrace{2N_t}^\text{target} + \overbrace{N_c + 4 \begin{cases}1 & \text{if } N_c = 3 \\ N_c & \text{if } N_c >3\end{cases}}^\text{comb} + \overbrace{N_c(3N_t)}^\text{interaction}\quad \stackrel{\mathlarger\sim}{\scalebox{0.8}{\tiny\text{large N}}} \quad N_cN_t
\end{equation}
\end{widetext}
where all $\textit{SWAP}$ gates have been ignored under the assumption that the quantum hardware will have all-to-all connectivity.  The role of all $\textit{SWAP}$ gates in the above circuitry are then solely for notational convenience and contribute nothing to the simulation's expense.
\par As long as the definition of $\mathbf{A}$ as the one-body part of the target Hamiltonian is sufficient to couple the states of the target to those of the comb, the additional resources  needed to couple a comb to any target system (not including those to simulate the target itself) will scale as $N_cN_t$.  As is true also for the interaction propagator circuit and all but $N_\text{c}$ gates for the comb (from the time-dependent one-body part of Eq.~\eqref{eq:Hcomb}), this entire circuit can be consolidated into a single unitary operator: the time evolution of the whole system may be summarized as $1$ unitary operator and $N_c$ single-qubit operators that change with time.  The interest of this statement comes in the possibility that there exists a more efficient implementation or suitable approximation to the time-independent propagator using an alternate elementary gate set.

\end{document}